\documentclass[epj]{webofc}
\usepackage[varg]{txfonts}   
\woctitle{International Conference on New Frontiers in Physics 2013}
\begin{document}
\title{Unconstrained Hamiltonian formulation of low energy QCD}
%
%

\author{Hans-Peter Pavel\inst{1,2}\fnsep\thanks{\email{hans-peter.pavel@physik.tu-darmstadt.de}}}

\institute{Bogoliubov Laboratory of Theoretical Physics, Joint Institute for Nuclear Research, Dubna, Russia  
\and
            Institut f\"ur Kernphysik (Theoriezentrum), TU Darmstadt, D-64289 Darmstadt, Germany
 }

\abstract{%
Using a generalized polar decomposition of the gauge fields into gauge-rotation and gauge-invariant parts,
which Abelianises the Non-Abelian Gauss-law constraints, an unconstrained Hamiltonian formulation of QCD
can be achieved. The exact implementation of the Gauss laws reduces the colored spin-1 gluons and spin-1/2 quarks
to unconstrained colorless spin-0, spin-1, spin-2 and spin-3 glueball fields and colorless Rarita-Schwinger fields.
The obtained physical Hamiltonian naturally admits a systematic strong-coupling expansion
in powers of $\lambda=g^{-2/3}$, equivalent to an expansion in the number of spatial derivatives.
The leading-order term corresponds to non-interacting hybrid-glueballs, whose low-lying spectrum can be calculated
with high accuracy by solving the Schr\"odinger-equation of the Dirac-Yang-Mills quantum mechanics of spatially
constant fields (at the moment only for the 2-color case). The discrete glueball excitation spectrum shows a
universal string-like behaviour with practically all excitation energy going in to the increase of the strengths
of merely two fields, the "constant Abelian fields" corresponding to the zero-energy valleys of the chromomagnetic
potential. Inclusion of the fermionic degrees of freedom significantly lowers the spectrum and allows for the study
of the sigma meson. Higher-order terms in $\lambda$ lead to interactions between the hybrid-glueballs
and can be taken into account systematically using perturbation theory, allowing for the study of IR-renormalisation
and Lorentz invariance. The existence of the generalized polar decomposition used, the position of the zeros
of the corresponding Jacobian (Gribov horizons), and the ranges of the physical variables can be investigated
by solving a system of algebraic equations. It is exactly solvable for 1 spatial dimension and several numerical
solutions can be found for 2 and 3 spatial dimensions.
}
\maketitle
%

\section{Introduction}
\label{intro}

The QCD action
\begin{eqnarray}
{\cal S} [A, \psi,\overline{\psi}] &  = &
\int d^4x\left[ - \frac{1}{4} F^a_{\mu\nu} F^{a\mu \nu}+\overline{\psi}\left(i\gamma^\mu D_\mu-m\right) \psi\right]
\end{eqnarray}
is invariant under the SU(3) gauge transformations
$U[\omega(x)]\equiv \exp(i\omega_a\tau_a/2)$
\begin{eqnarray}
\label{tr}
\psi^{\omega}(x) = U[\omega(x)]\ \psi(x),\quad\quad
A_{a\mu}^{\omega}(x) \tau_a/2   =
U[\omega(x)] \left(A_{a\mu}(x) \tau_a/2 +{i\over g}\partial_\mu \right) U^{-1}[\omega(x)].
\end{eqnarray}
Introducing the chromoelectric
$E_{i}^{a}\equiv F^a_{i 0}$ and chromomagnetic  $B_{i}^{a} \equiv\frac{1}{2}\epsilon_{ijk}F^a_{j k}$  and noting that 
the momenta conjugate to the spatial $A_{ai}$ are $\Pi_{ai}=-E_{ai}$, we obtain the canonical Hamiltonian
\begin{eqnarray}
\!\! H_C \!\! &=& \!\!\int d^3x\Bigg[{1\over 2}E_{ai}^2+{1\over 2} B_{ai}^2(A) - g A_{ai}\, j_{i a}(\psi)
   + \overline{\psi}\left(\gamma_i\partial_i+m\right) \psi
   - g A_{a0} \left( D_i(A)_{ab} E_{bi} - \rho_a(\psi)\right)\Bigg],
\end{eqnarray}
with the covariant derivative
$~D_i(A)_{ab}\equiv\delta_{ab}\partial_i-g f_{abc}A_{ci}~$ in the adjoint representation.

Exploiting the {\it time dependence of the gauge transformations} (\ref{tr}) to put (see e.g. \cite{Christ and Lee})
\begin{equation}
A_{a0} = 0~,\quad\quad a=1,..,8 \quad\quad ({\rm Weyl\ gauge}),
\end{equation}
and quantising the dynamical variables $A_{ai}$, $-E_{ai}$, $\psi_{\alpha r}$ and $\psi^*_{\alpha r}$
in the Schr\"odinger functional approach
 by imposing equal-time (anti-) commutation relations (CR) , e.g.  $-E_{ai} = -i\partial/\partial A_{ai}$,
the physical states $\Phi$ have to satisfy both the Schr\"odinger equation and the Gauss laws 
\begin{eqnarray}
\label{constrH}
H\Phi &=& \int d^3x
 \left[{1\over 2} E_{ai}^2+{1\over 2}B_{ai}^2[A]
- A_{ai}\, j_{i a}(\psi)+ \overline{\psi}\left(\gamma_i\partial_i+m\right) \psi\right]\Phi=E\Phi~,
\quad\quad \\
G_a(x)\Phi &=& \left[D_i(A)_{ab} E_{bi}
-\rho_a(\psi)\right]\Phi=0~,
\quad a=1,..,8~.
\label{constrG}
\end{eqnarray}
The Gauss law operators $G_a$ are the generators of the residual {\it time independent gauge transformations} in (\ref{tr}).
They satisfy $[G_a(x),H]=0$ and $[G_a(x),G_b(y)]=if_{abc}G_c(x)\delta(x-y)$.

\noindent
Furthermore, $H$ commutes with the angular momentum operators  
\begin{eqnarray}
\label{constrainedJ}
 J_i  = \int d^3x\left[  -\epsilon_{ijk}A_{aj} E_{ak} + \Sigma_i(\psi)+ {\rm orbital \ parts}\right] ~,
\quad i=1,2,3~.
\end{eqnarray}
The matrix element of an operator $O$ is given in the {\it Cartesian} form
\begin{eqnarray}
\langle \Phi'| O|\Phi\rangle\
\propto
\int dA\ d\overline{\psi}\ d\psi\
\Phi'^*(A,\overline{\psi},\psi)\, O\, \Phi(A,\overline{\psi},\psi)~.
\end{eqnarray}

For the case of Yang-Mills quantum mechanics of spatially constant gluon fields, 
the  spectrum of Equ.(\ref{constrH})-(\ref{constrG}) has been found in \cite{Luescher and Muenster} for SU(2) 
and in \cite{Weisz} for SU(3), in the context of a weak coupling expansion in $g^{2/3}$, using the variational approach with gauge-invariant wave-functionals automatically satisfying (\ref{constrG}). 
The corresponding unconstrained approach, 
a description in terms of gauge-invariant dynamical variables via an exact implementation of the Gaws laws, 
has been considered by many authors
(o.a. \cite{Christ and Lee},\cite{Kihlberg and Marnelius}-\cite{pavel2013}, and references therein) 
to obtain a non-perturbative description of QCD at low energy, as an alternative to lattice QCD. 

I shall first discuss in Sec.\ref{sec-2} the unphysical, but technically much simpler case of 2-colors, and then show in Sec.\ref{sec-3}
and Sec.\ref{sec-4} how the results can be generalised to SU(3).

\section{Review: Unconstrained Hamiltonian formulation of 2-color QCD}
\label{sec-2}

\subsection{Canonical transformation to adapted coordinates}

Point transformation from the  $A_{ai},\psi_\alpha$ to a new set of adapted coordinates,
the  3 angles $ q_j$ of an orthogonal matrix $O(q)$, 
the  6 elements of a pos. definite symmetric $3\times 3$ matrix $S$, and  new  $\psi^\prime_\beta$
\begin{eqnarray}
\label{polardecomp}
A_{ai} \left(q, S \right) =
O_{ak}\left(q\right) S_{ki}
- {1\over 2g}\epsilon_{abc} \left( O\left(q\right)
\partial_i O^T\left(q\right)\right)_{bc}\,,\quad
\psi_\alpha\left(q, \psi^\prime \right)= U_{\alpha\beta}\left( q \right) \psi^\prime_{\beta}~, 
\end{eqnarray}
where the orthogonal  $ O(q) $  and the unitary   $U\left( q \right)$ are related  via
$O_{ab}(q)={1\over 2}\mbox{Tr}\left(U^{-1}(q)\tau_a U(q)\tau_b\right)$.
Equ. (\ref{polardecomp}) is the generalisation of the (unique) polar decomposition of $A$ and corresponds to 
\begin{eqnarray}
\label{symgaugeSU2}
\quad\quad\chi_i(A)=\epsilon_{ijk}A_{jk}=0\quad(\rm{"symmetric}\ \rm{gauge"}).
\end{eqnarray}
Preserving the CR, we obtain the old canonical momenta in terms of the new variables
\begin{eqnarray}
-E_{ai}(q,S,p,P)=O_{ak}\left(q\right)\left[ P_{ki}+\epsilon_{kil}
{^{\ast}\! D}^{-1}_{ls}(S)\left(\Omega^{-1}_{sj}(q) p_{j}
+\rho_s(\psi^{\prime})
+D_n(S)_{sm} P_{mn}\right)\right].
\end{eqnarray}
In terms of the new canonical variables the Gauss law constraints are Abelianised,
\begin{eqnarray}
\label{unconstrG}
  G_a\Phi
\equiv  O_{ak}(q)\Omega^{-1}_{ki}(q)  p_i\Phi =0\quad
\Leftrightarrow\quad\frac{\delta }{\delta q_i}\Phi=0\quad (\rm{Abelianisation}),
\end{eqnarray}
and the angular momenta become
\begin{eqnarray}
\label{unconstrJ}
 J_i  = \int d^3x\left[  -2\epsilon_{ijk}S_{mj} P_{mk} + 
\Sigma_i(\psi^{\prime})+ \rho_i(\psi^{\prime})+ {\rm orbital \ parts}\right].
\end{eqnarray}
Equ.(\ref{unconstrG}) identifies the $q_i$ with the gauge angles and $S$ and $\psi^{\prime}$ as the physical fields.
Furthermore, from Equ.(\ref{unconstrJ}) follows that the $S$ are {\it colorless} 
spin-0 and spin-2 glueball fields, and the $\psi^{\prime}$ {\it colorless} reduced quark fields of spin-0 and spin-1.
Hence the gauge reduction corresponds to the conversion {\it "color $\rightarrow$ spin"}. The obtained unusual
spin-statistics relation is specific to SU(2).
 
\subsection{Physical quantum Hamiltonian}

The {\it correctly ordered physical quantum Hamiltonian} in terms of the
physical variables $S_{ik}({\mathbf{x}})$  and the canonically conjugate 
$P_{ik}({\mathbf{x}})\equiv -i\delta /\delta S_{ik}({\mathbf{x}})$ reads \cite{pavel2010} (following the general scheme \cite{Christ and Lee}) ,
\begin{eqnarray}
H(S,P)\!\!\!\!&=&\!\!\!\! {1\over 2}{\cal J}^{-1}\!\!\int d^3{\mathbf{x}}\ P_{ai}\ {\cal J}P_{ai}
     +{1\over 2}\int d^3{\mathbf{x}}\left[B_{ai}^2(S)- S_{ai}\, j_{i a}(\psi^\prime) + \overline{\psi}^\prime\left(\gamma_i\partial_i+m\right) \psi^\prime\right]\nonumber\\
      & & -{\cal J}^{-1}\!\!\int d^3{\mathbf{x}}\int d^3{\mathbf{y}}
      \Big\{\Big(D_i(S)_{ma}P_{im}+\rho_a(\psi^\prime)\Big)({\mathbf{x}}){\cal J}\
        \nonumber\\
      &&\quad\quad\quad\quad\quad\quad\quad\quad\quad\quad
            \langle {\mathbf{x}}\ a|^{\ast}\! D^{-2}(S)|{\mathbf{y}}\ b\rangle\ \
             \Big(D_j(S)_{bn}P_{nj}+\rho_b(\psi^\prime)\Big)({\mathbf{y}})\Big\},
\end{eqnarray}
with the Faddeev-Popov (FP) operator 
\begin{eqnarray}
^{\ast}\! D_{kl}(S)\equiv \epsilon_{kmi}D_i(S)_{ml}=\epsilon_{kli}\partial_i-g (S_{kl}-\delta_{kl} {\rm tr} S),
\end{eqnarray}
and the Jacobian
${\cal J}\equiv\det |^{\ast}\! D| $.
The matrix element of a physical operator O is given by
\begin{eqnarray}
\langle \Psi'| O|\Psi\rangle\
\propto
\int_{\rm S\ pos. def.}\int_{\overline{\psi}^\prime\!,\psi^\prime } 
 \prod_{\mathbf{x}}\Big[dS({\mathbf{x}})d \overline{\psi}^\prime\! ({\mathbf{x}})d \psi^\prime\!( {\mathbf{x}})\Big]
{\cal J}\Psi'^*[S,\overline{\psi}^\prime,\psi^\prime] O\Psi[S,\overline{\psi}^\prime,\psi^\prime].
\end{eqnarray}
 {\it The inverse of the FP operator and hence the physical Hamiltonian
can be expanded in the number of spatial derivatives, equivalent to a strong coupling expansion in $\lambda=g^{-2/3}$}.

\subsection{Coarse-graining and strong coupling expansion of the physical Hamiltonian in $\lambda\!=\!g^{-2/3}$}

Introducing an UV cutoff $a$ by considering an infinite spatial lattice of granulas
$G({\mathbf{n}},a)$ at  ${\mathbf{x}}=a {\mathbf{n}}$  $({\mathbf{n}}\in Z^3)$
and averaged variables
\begin{eqnarray}
\label{average}
S({\mathbf{n}}) :=  \frac{1}{a^3}\int_{G({\mathbf{n}},a)} d{\mathbf{x}}\ S({\mathbf{x}})~,
\end{eqnarray}
and discretised spatial derivatives, the expansion of the Hamiltonian in $\lambda=g^{-2/3}$ can be written
\begin{eqnarray}
H =\frac{g^{2/3}}{a}\left[{\cal H}_0+\lambda \sum_\alpha {\cal V}^{(\partial)}_\alpha
                             +\lambda^2 \left(\sum_\beta {\cal V}^{(\Delta)}_\beta
                             +\sum_\gamma{\cal V}^{(\partial\partial\neq\Delta)}_\gamma\right)
                             + {\mathcal{O}}(\lambda^3)\right].
\end{eqnarray}
The "free" Hamiltonian
$ H_0 = (g^{2/3}/a){\cal H}_0+{H}_m
=\sum_{\mathbf{n}} H_0^{QM}({\mathbf{n}}) $
is the sum of the Hamiltonians of Dirac-Yang-Mills quantum mechanics of constant
fields in each box, and the interaction terms 
$ {\cal V}^{(\partial)}, {\cal V}^{(\Delta)},..$ lead to interactions between the granulas.

\subsection{Zeroth-order: Dirac-Yang-Mills quantum mechanics of spatially constant fields}

\noindent
Transforming to the intrinsic system of the symmetric tensor $S$, with Jacobian $\sin\beta\prod_{i<j}(\phi_i-\phi_j)$,
\begin{eqnarray}
S  =  R^{T}(\alpha,\beta,\gamma)\ \mbox{diag} ( \phi_1 , \phi_2 , \phi_3 )\ R(\alpha,\beta,\gamma),
\quad\quad \psi^{\prime (i)}_{L,R}=R^T_{ij}\widetilde{\psi}^{ (j)}_{L,R},
\quad\quad \psi^{\prime (0)}_{L,R}=\widetilde{\psi}^{ (0)}_{L,R},
\end{eqnarray}
the "free"  Hamiltonian in each box of volume $V$ takes the form \cite{pavel2011}
\begin{eqnarray}
H_0^{QM}&=&{g^{2/3}\over V^{1/3}}\Bigg[{\cal H}^{G}+{\cal H}^{D}+{\cal H}^{C}\Bigg]
+{1\over 2} m \left[\left(\widetilde{\psi}^{(0)\dag}_L\widetilde{\psi}^{(0)}_R\!
         +\sum_{i=1}^3\widetilde{\psi}^{(i)\dag}_L\widetilde{\psi}^{(i)}_R\right)\! +h.c.\right],
\end{eqnarray}
with the glueball part  ${\cal H}^{G}$, the minimal-coupling ${\cal H}^{D}$, and the Coulomb-potential-type part ${\cal H}^{C}$
\begin{eqnarray}
 {\cal H}^{G}\!\!&=&\!{1\over 2}\!\sum^{\rm
cyclic}_{ijk} \Bigg(\! -{\partial^2\over\partial \phi_i^2} -{2\over \phi_i^2-
\phi_j^2}\left(\phi_i{\partial\over\partial
\phi_i}-\phi_j{\partial\over\partial \phi_j}\right)
 +(\xi_i-\widetilde{J}^Q_i)^2  {\phi_j^2+\phi_k^2\over (\phi_j^2-\phi_k^2)^2}
+  \phi_j^2 \phi_k^2\Bigg) ,
\\
\label{HphysD}
{\cal H}^{D}\!\!&=&\!\! {1\over 2}(\phi_1+\phi_2+\phi_3)
\left(\widetilde{N}^{(0)}_L-\widetilde{N}^{(0)}_R\right)
+
{1\over 2}\sum^{\rm cyclic}_{ijk} (\phi_i-(\phi_j+\phi_k))
\left(\widetilde{N}^{(i)}_L-\widetilde{N}^{(i)}_R\right),
\\
\label{HphysC}
{\cal H}^{C}\!\!&=&
\!\! \sum^{\rm cyclic}_{ijk}
      {\widetilde{\rho}_i(\xi_i-\widetilde{J}^Q_i+\widetilde{\rho}_i) \over (\phi_j+\phi_k)^2}~,
\end{eqnarray}
\vspace{-0.5cm}
\begin{eqnarray}
{\rm and\ the\ total\ spin }\quad\quad\quad\quad J_i=R_{ij}(\chi)\ \xi_j~,\quad\quad [J_i,H]=0~.
\quad\quad\quad\quad\quad\quad\quad\quad\quad\quad\quad\quad\quad\quad\ \ 
\end{eqnarray}
The matrix elements become
\begin{eqnarray}
\langle\Phi_1|{\cal O} |\Phi_2\rangle \!=\!
\!\!\int\!\!  d\alpha\, \sin\beta\, d\beta\, d\gamma\!\! \int_{0<\phi_1<\phi_2<\phi_3}
\!\!\!\!\!\!\!\!\!\!\!\!\!\!\!\!\!\!\!\!\!
d\phi_1d\phi_2d\phi_3\ 
(\phi_1^2\!-\!\phi_2^2)(\phi_2^2\!-\!\phi_3^2)(\phi_3^2\!-\!\phi_1^2)\int d\overline{\psi}^\prime d\psi^\prime\Phi_1^* {\cal O}\Phi_2~.\nonumber
\end{eqnarray}

The l.h.s. of Fig.1 shows the $0^+$ energy spectrum of the lowest pure-gluon (G) and quark-gluon (QG) cases for one quark-flavor   
 which can be calculated with high accuracy using the variational approach. 
The energies of the quark-gluon ground state and the sigma-antisigma excitation
are lower than that of the lowest pure-gluon state. This is due to a large negative contribution from 
$\langle{\cal H}^{D}\rangle$, in addition to the large positive $\langle{\cal H}^{G}\rangle$,
while  $\langle{\cal H}^{C}\rangle\simeq 0\ $ (see \cite{pavel2011} for details).

Furthermore, as a consequence of the zero-energy valleys $\ "\phi_1\!=\!\phi_2\!=0,\ \ \phi_3\  {\rm arbitrary}"$ of the classical  magnetic  potential $B^2=\phi_2^2\phi_3^2+\phi_3^2\phi_1^2+\phi_1^2\phi_2^2 $, 
practically all glueball excitation-energy results from an increase of expectation value
of the "constant Abelian field" $\phi_3$ as shown for the pure-gluon case on the r.h.s. of Fig.1 (see \cite{pavel2007} for details). 

\begin{figure}
\includegraphics[width=0.55\textwidth]{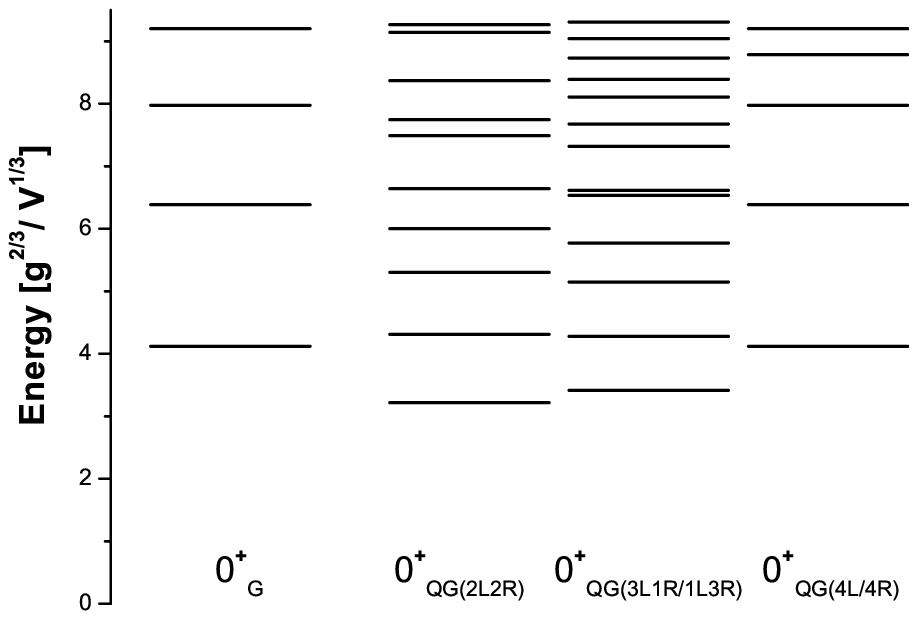}
\includegraphics[width=0.45\textwidth]{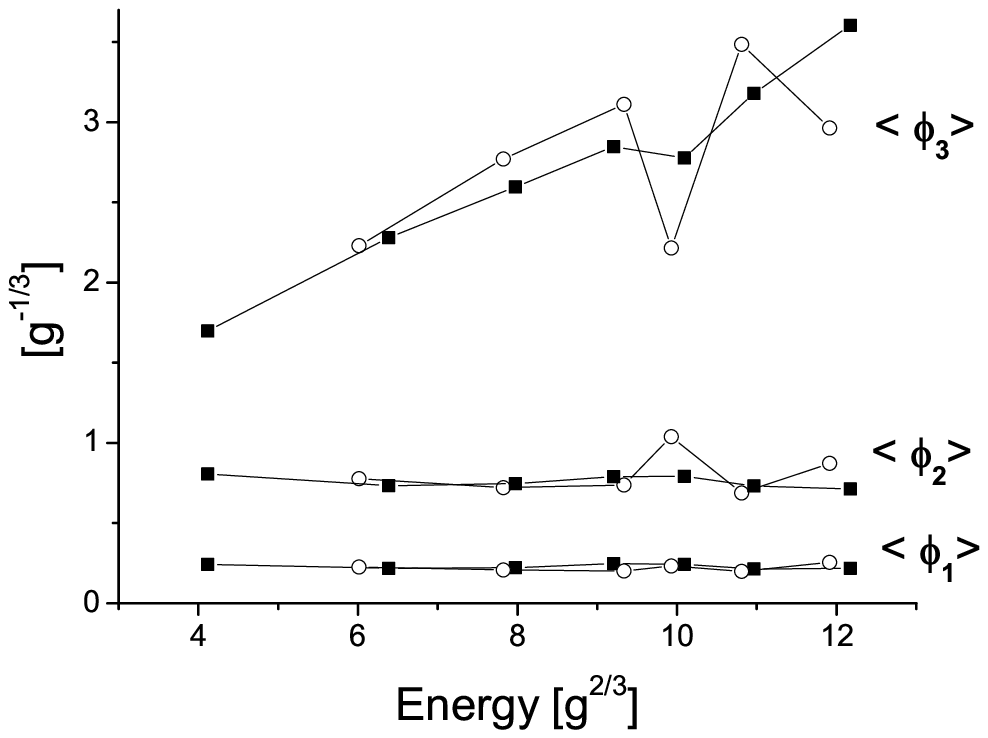}
\caption{L.h.s.: Lowest energy levels for the pure-gluon (G) and the quark-gluon case (QG) for 2-colors and one quark flavor. 
The energies of the quark-gluon ground state and the sigma-antisigma excitation
are lower than that of the lowest pure-gluon state. R.h.s. (for pure-gluon case and setting $V\equiv 1$): 
$\langle \phi_3\rangle$ is raising with increasing excitation,
whereas $\langle \phi_1\rangle$ and $\langle \phi_2\rangle$ are practically constant,
independent of whether spin-0 (dark boxes) or spin-2 states (open circles).}
\label{fig1}
\end{figure}

\subsection{Perturbation theory in $\lambda$ and coupling constant renormalisation in the IR }

Inclusion of the interactions ${\cal V}^{(\partial)}, {\cal V}^{(\Delta)}$ using 1st and 2nd order perturbation theory in  $\lambda=g^{-2/3}$ 
give the result \cite{pavel2010} (for the pure-gluon case and including only spin-0 fields in a first approximation)
\begin{eqnarray}
E_{\rm vac}^{+}\!&=&\!{\cal N}\frac{g^{2/3}}{a}\Bigg[4.1167+29.894\lambda^2+{\cal O}(\lambda^3)\Bigg],\\
 E_1^{(0)+}(k)-E_{\rm vac}^{+}\! &=&\!
 \left[\ 2.270 + 13.511\lambda^2 + {\cal O}(\lambda^3)\right]\frac{g^{2/3}}{a}
 +0.488 \frac{a}{g^{2/3}} k^2 +{\cal O}((a^2 k^2)^2) ,
\end{eqnarray}
for the energy of the interacting glueball vacuum and the spectrum of the interacting spin-0 glueball.
Lorentz invariance demands $E=\sqrt{M^2+k^2}\simeq M + \frac{1}{2 M}\ k^2 $, which is violated in this 1st approximation
by a factor of $2$. In order to get a Lorentz invariant result,  $J=L+S$ states should be considered including also spin-2 states 
and the general ${\cal V}^{(\partial\partial)}$.

Demanding independence of the physical glueball mass
\begin{eqnarray}
M = \frac{g_0^{2/3}}{a}\left[\mu +c g_0^{-4/3}\right]\nonumber
\end{eqnarray}
of box size $a$, one obtains
\begin{eqnarray}
\gamma(g_0) \equiv a {d\over da} g_0(a) = \frac{3}{2} g_0\frac{\mu+c g_0^{-4/3}} {\mu-c g_0^{-4/3}}~,
\end{eqnarray}
which  vanishes for 
$g_0 = 0$ (pert.\ fixed\ point) or $g_0^{4/3}=-c/\mu$ (IR\ fixed\ point, if $c<0$). For $c>0$ 
\begin{eqnarray}
{\rm for}\ c>0:\quad\quad g_0^{2/3}(M a) =\frac{M a}{2\mu}
     +\sqrt{\left(\frac{M a}{2\mu}\right)^2
     -\frac{c}{\mu}}~,\quad\quad
     \  a >  a_{c} :=  2\sqrt{c \mu}/M
\end{eqnarray}
My (incomplete) result  $ c_1^{(0)}/\mu_1^{(0)}=5.95 $ suggests, that no IR fixed points exist. For the
critical coupling I find $g^2_{0}|_{c}= 14.52$ and $ a_{c} \sim 1.4\ {\rm fm}$ for $ M \sim 1.6\ {\rm GeV}$.

\section{Symmetric gauge for SU(3)}
\label{sec-3}

Using the idea of  {\it minimal embedding} of $su(2)$ in $su(3)$ by Kihlberg and Marnelius \cite{Kihlberg and Marnelius}
\begin{eqnarray}
\label{minimal embedding}
&&\!\!\!\!\!\!\!\!\!\!\!\!
\tau_1 := \lambda_7 ={\small \left(\begin{array}{c c c} 0&0&0\\ 0&0&-i\\ 0&i&0  \end{array}\right)}
\quad
\tau_2 := -\lambda_5 ={\small  \left(\begin{array}{c c c} 0&0&i\\ 0&0&0\\ -i&0&0  \end{array}\right)}
\quad
\tau_3 := \lambda_2 =\!{\small  \left(\!\!\begin{array}{c c c} 0&-i&0\\ i&0&0\\ 0&0&0  \end{array}\!\!\right)}
\nonumber\\
&&\!\!\!\!\!\!\!\!\!\!\!\!
\tau_4 := \lambda_6 = {\small \left(\begin{array}{c c c} 0&0&0\\ 0&0&1\\ 0&1&0  \end{array}\right)}
\quad\quad\!\!
\tau_5 := \lambda_4 = {\small \left(\begin{array}{c c c} 0&0&1\\ 0&0&0\\ 1&0&0  \end{array}\right)}
\quad\quad\
\tau_6 := \lambda_1 =\! {\small \left(\!\!\begin{array}{c c c} 0&1&0\\ 1&0&0\\ 0&0&0  \end{array}\!\!\right)}
\nonumber\\
&&\!\!\!\!\!\!\!\!\!\!\!\!
\tau_7 := \lambda_3 = {\small \left(\begin{array}{c c c} 1&0&0\\ 0&-1&0\\ 0&0&0  \end{array}\right)}
\quad
\tau_8 := \lambda_8 = {1\over \sqrt{3}}{\small \left(\begin{array}{c c c} 1&0&0\\ 0&1&0\\ 0&0&-2  \end{array}\right)}
\end{eqnarray}
such that the corresponding non-vanishing structure constants, 
$[{\tau_a\over 2},{\tau_b\over 2}]=  i c_{abc} {\tau_c\over 2}$, have at least one index  $\in \{1,2,3\}$, 
the symmetric gauge of SU(2), Equ.(\ref{symgaugeSU2}), can be generalised to SU(3) \cite{Dahmen and Raabe, pavel2012,pavel2013},
\begin{eqnarray}
\chi_a(A)=\sum_{b=1}^8\sum_{i=1}^3 c_{abi} A_{bi}=0~,\quad a=1,...,8\quad\quad ({\rm "symmetric\ gauge"\  for\ SU(3)}).
\end{eqnarray}
Carrying out the coordinate transformation \cite{pavel2012}
\begin{eqnarray}
A_{ak}\!\left(q_1,..,q_8, \widehat{S} \right)\!
&=&\! O_{a\hat{a}}\left(q \right) \widehat{S} _{\hat{a} k}- {1\over 2g}c_{abc} \left( O\left(q\right)
\partial_k O^T\!\!\left(q\right)\right)_{bc},
\quad
\psi_\alpha\left(q_1,..,q_8, \psi^{RS} \right)= U_{\alpha{\hat{\beta}}}\left( q \right) \psi^{RS}_{\hat{\beta}}\nonumber\\
 \widehat{S}_{\hat{a} k }
\equiv {S_{i k }\choose \overline{S}_{A k} }
&=&
\left(\begin{array}{c c c}
                  &  &  \\
 & S_{ik} \ {\rm pos.\ def.}                  & \\
 & & 
\\ \hline
W_0 &
X_3 -W_{3}&
X_2  +W_{2}
\\
X_3  +W_{3}&
W_0 &
X_1  -W_{1}
\\
X_2  -W_{2}&
X_1  +W_{1}&
W_0
\\  
 -{\sqrt{3}\over 2}Y_1  -{1\over 2}W_{1}&
 {\sqrt{3}\over 2}Y_2   -{1\over 2}W_{2}&
  W_{3}
\\
  - {\sqrt{3}\over 2}W_{1} -{1\over 2}Y_1&
  {\sqrt{3}\over 2}W_{2} -{1\over 2}Y_2&
 Y_3
\end{array}\right),\quad  c_{\hat{a}\hat{b} k} \widehat{S}_{\hat{b} k }=0~,
\label{unconstrSU3}
\end{eqnarray}
an unconstrained Hamiltonian formulation of QCD can be obtained.  
The existence and uniqueness of (\ref{unconstrSU3}) can be investigated by
solving the 16 equs.
\begin{eqnarray}
\label{SAequs}
\quad \widehat{S}_{\hat{a} i }\widehat{S}_{\hat{a} j }= A_{ai}A_{aj}\ (6\ {\rm equs.}) 
\quad\wedge\quad 
d_{\hat{a}\hat{b}\hat{c}}\widehat{S}_{\hat{a} i }\widehat{S}_{\hat{b} j }\widehat{S}_{\hat{c} k }= d_{abc}A_{ai}A_{bj}A_{ck}\ (10 \ {\rm equs.}) 
\end{eqnarray}
for the 16 components of $ \widehat{S}$ in terms of 24 given components A, and will be further discussed in Sec.\ref{sec-4}.

Expressing the Gauss law operators and the unconstrained angular momentum
operators in terms of the new variables in analogy to the 2-color case, it can be shown that the original constrained 24 colored spin-1 gluon fields $A$
and the 12 colored spin-1/2 quark fields $\psi$ (per flavor)
reduce to {\it 16 physical  colorless spin-0, spin-1, spin-2, and spin-3 glueball fields} (the 16 components of $\widehat{S}$) 
and {\it a colorless spin-3/2 Rarita-Schwinger field} $\psi^{RS}$ (12 components per flavor).
As for the 2-color case, the gauge reduction converts  {\it color $\rightarrow$ spin}, which might have important
consequences for low energy Spin-Physics. In terms of the colorless Rarita-Schwinger fields the  
$\Delta^{++}(3/2)$ could have the spin content $(+3/2,+1/2,-1/2)$ in accordance with the Spin-Statistics-Theorem.

\noindent
Transforming to the intrinsic system of the embedded upper part S of $ \widehat{S}$  (see \cite{pavel2012} for details)
\begin{eqnarray}
S  =  R^{T}(\alpha,\beta,\gamma)\ \mbox{diag} ( \phi_1 , \phi_2 , \phi_3 )\  R(\alpha,\beta,\gamma),
\ \ \wedge\ \ X_i\rightarrow  x_i,\ Y_i\rightarrow y_i ,..,\ \ \wedge\ \ {\psi}^{RS}\rightarrow \widetilde{\psi}^{RS},
\end{eqnarray}
one finds that the magnetic potential $B^2$  has the zero-energy valleys ("constant Abelian fields")
\begin{eqnarray}
B^2=0\ \ :\ \ \phi_3\ {\rm and}\  y_3\  {\rm  arbitrary }\quad \wedge \quad  {\rm all\ others\ zero}
\end{eqnarray}
Hence, practically all glueball excitation-energy should result from an increase of expectation values
of these two "constant Abelian fields", in analogy to SU(2) . 
Furthermore, at the bottom of the valleys the important minimal-coupling-interaction of $\widetilde{\psi}^{RS}$ 
(analogous to (\ref{HphysD})) becomes diagonal
\begin{eqnarray}
 {\cal H}^D_{\rm diag}=
{1\over 2}\widetilde{\psi}_L^{(1,{1\over 2})\dagger}
\left[\left( \phi_3 \lambda_3+y_3 \lambda_8 \right)\otimes \sigma_3 \right]\widetilde{\psi}^{(1,{1\over 2})}_L
 -  {1\over 2}\widetilde{\psi}_R^{({1\over 2},1)\dagger}
\left[ \sigma_3\otimes\left( \phi_3 \lambda_3+y_3 \lambda_8 \right) \right]\widetilde{\psi}^{({1\over 2},1)}_R~.
\end{eqnarray}
Due to the difficulty of the FP-determinant (see \cite{pavel2012}), precise calculations are not yet possible.

\section{Symmetric gauge for SU(3) in one, two and three spatial dimensions}
\label{sec-4}

\subsection{One spatial dimension}
\label{sec-1dim}

\begin{figure}
\centering
\includegraphics[width=5cm,clip]{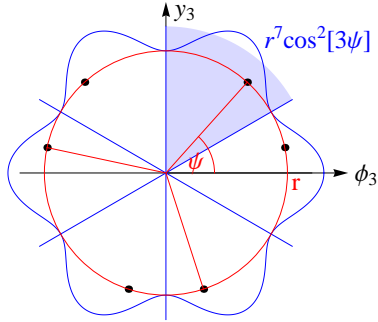}
\caption{In each of the six Weyl-chambers, separated by the radial lines $\psi_k=(1+2k)\pi/6$ of vanishing Jacobian ("Gribov-horizons"), lies one solution (black dots) of Eqs. (\ref{SAequs2dim}) for a given 1-dim $A^{(1d)}$. As fundamental domain one can e.g. choose the shaded area.}
\label{fig-2}  
\end{figure}

In one spatial dimension the symmetric gauge for SU(3) reduces to the t'Hooft gauge\footnote{%
The $A_3$ field is diagonalised in the fundamental representation, $ A_3^{a}\lambda_{a}\!=\!U (\phi_3\lambda_{3}\!+\!y_3\lambda_{8}) U^+)$}
\begin{eqnarray}
{ A^{(1d)}=
\left(\begin{array}{c c c}
0     & 0 &A_{13}  \\
0 &   0   &A_{23} \\
0 &0 &A_{33} \\ 
0&0&A_{43}\\
0&0&A_{53}\\  
0&0&A_{63}\\
0&0&A_{73}\\
0&0&A_{83} 
\end{array}\right)
\quad\rightarrow\quad S^{(1d)}=
\left(\begin{array}{c c c}
0     & 0 &0  \\
0 &   0   &0 \\
0 &0 & \phi_3\\ 
\hline
0&0&0\\
0&0&0\\  
0&0&0\\
0&0&0\\
0&0& y_3
\end{array}\right)~,
\nonumber
}\end{eqnarray}
which consistently reduces the above Eqs.(\ref{SAequs}) for $S$ for given $A_3$ to
\begin{eqnarray}
 \phi_3^2+y_3^2=A_{a3}A_{a3}\quad\wedge \quad \phi_3^2\, y_3-3\, y_3^3 = d_{abc}A_{a3}A_{b3}A_{c3}  ~.
\label{SAequs2dim}
\end{eqnarray}
In terms of polar coordinates $\phi_3=r \cos\psi$ and $y_3=r \sin\psi$ Eqs.(\ref{SAequs2dim}) read
\begin{eqnarray}
 r^2=A_{a3}A_{a3}\quad\wedge \quad r^3 \sin(3\psi) =\sqrt{3}\, d_{abc}A_{a3}A_{b3}A_{c3} ~, 
\nonumber
\end{eqnarray}
with 6 solutions, see black dots in Fig.\ref{fig-2} , one in each Weyl-chamber,  separated by the zero-lines  of the FP-determinant
\begin{eqnarray}
 r^7\cos^2(3\psi) =0   \quad\quad {\rm for}   \quad\quad \psi_k=(1+2k)\frac{\pi}{6}      \quad\quad   {\rm ( "Gribov-horizons")}~.
\nonumber
\end{eqnarray}
Exactly one solution exists in the  "fundamental domain" (shaded area in Fig.\ref{fig-2})
and we can replace
\begin{eqnarray}
\int_{-\infty}^{+\infty} \prod_{a=1}^8 d A_{a3}
&\rightarrow &\int_0^\infty\!\!\!\!\!\! d \phi_3\! \int_{ \phi_3/\sqrt{3}}^\infty d y_3\  \phi_3^2\left( \phi_3^2-3 y_3^2\right)^2
\propto\int_0^\infty\!\!\! r^7 dr\! \int_{\pi/6}^{\pi/2}d \psi \cos^2(3\psi)~.
\nonumber
\end{eqnarray}

\subsection{Two spatial dimensions}

For two spatial dimensions, one can show that (putting in Equ.(\ref{unconstrSU3}) $W_1\equiv X_1,W_2\equiv -X_2$)
\begin{eqnarray}
A^{(2d)}=
{\left(\begin{array}{c c c}
A_{11}  &A_{12} & 0 \\
 A_{21}    &A_{22} & 0  \\
A_{31} &A_{32} & 0  \\ 
A_{41}&A_{42} & 0 \\
A_{51}&A_{52} & 0 \\  
A_{61}&A_{62} & 0 \\
A_{71}&A_{72} & 0 \\
A_{81}&A_{82} & 0  
\end{array}\right)}
\rightarrow \widehat{S}^{(2d)}_{\rm intrinsic}= \ 
{\left(\begin{array}{c c c}
\phi_1     & 0 &0  \\
0 &   \phi_2    &0 \\
0 &0 & 0\\ 
\hline
0& x_3&0\\
 x_3&0&0\\  
2 x_2&2 x_1&0\\
-{\sqrt{3}\over 2}y_1- {1\over 2}x_1&\ {\sqrt{3}\over 2}y_2+ {1\over 2}x_2 &\ 0\\
-{\sqrt{3}\over 2}y_1+ {1\over 2}x_1&\ -{\sqrt{3}\over 2}y_2+ {1\over 2}x_2&\ 0
\end{array}\right)}
\end{eqnarray}
consistently reduces the above equs. (\ref{SAequs}) for $S$ to a system of $7$ equs. for $8$ physical fields (incl. rot.-angle $\gamma$), which, adding as an 8th equ.  
$
 (d_{\hat{a}\hat{b}\hat{c}}\widehat{S}_{\hat{b} 1} \widehat{S}_{\hat{c} 2 })^2= (d_{abc}A_{b 1} A_{c 2 })^2~,
$
can be solved numerically for randomly generated $A^{(2d)}$, without loss of generality normalised to $r^2\equiv A^{(2d)}_{ai}A^{(2d)}_{ai}=1$.
Although, to the best of my knowlegde, nobody has ever succeeded in finding all solutions using computer-algebra (e.g. Groebner basis methods),
I have found several solutions with arbitrary high numerical accuracy in all generated cases, indicating that the generalised
polar decomposition exists for any 2-dim gauge configuration.
Tab.\ref{tab-1} gives one example of several (6) solutions ("Gribov copies") for one given randomly generated gauge field, 
written in terms of intrinsic coordinates
and rotation angle $\gamma$ to the intrinsic frame. In the 2-dim we can always achieve $0<\phi_1<\phi_2$ and $y_2>0$ for a suitable $\gamma$.
\begin{table}
\centering
\caption{Several 2-dim solutions (Gribov-copies) for one randomly generated $A^{(2d)}$, written in terms of intrinsic coordinates
and rotation angle $\gamma$ to the intrinsic frame.}
\label{tab-1}       
\begin{tabular}{llllllllll}
\hline
     & $\gamma$ & $\phi_1$ & $\phi_2$ & $x_1$ & $x_2$ & $x_3$ & $y_1$ & $y_2$ \\
\hline
1    &0.65 &0.24  &0.81  &-0.08 &0.21 &0.24 &0.33 &0.002 \\
\hline
2   &0.87  & 0.42 &0.82  &0.07 &0.07 &0.31 &-0.18 &0.01\\
\hline
3   &1.81  &0.40  &0.53  &-0.17 &-0.35 &0.17 &0.40 &0.03 \\
\hline
4   &4.06  &0.34  &0.44  &-0.24 &-0.02 &0.43 &0.20 &0.57 \\
\hline
5   &3.99  & 0.08 & 0.74 &-0.03 &0.17 &0.57 &-0.20 &0.02 \\
\hline
6   &3.71  &0.12  & 0.47 &0.30 &-0.13 &0.34 &-0.39 &0.49 \\
\hline
\end{tabular}
\end{table}
Note that we can represent the intrinsic variables also in terms  of spin-0, spin-1, spin-2 and spin-3 combinations
 $\phi^{(0)}\equiv (\phi_1+\phi_2)/\sqrt{2}$, $\phi^{(2)}\equiv (\phi_2-\phi_1)/\sqrt{2}$, 
$v_i^{(1)}\equiv\left(\sqrt{15}x_i-\sqrt{10}y_i\right)/4$, and $w_i^{(3)}\equiv\left(5x_i+\sqrt{6}y_i\right)/4$ for $i=1,2$. The variable $x_3$
turns out to be a scalar under 2-dim rotations.
Tab.\ref{tab-2} shows the spin-0, spin-1,spin-2, and spin-3 contents $\phi^{(0)}$, $v^{(1)}\equiv \sqrt{v_1^{(1)2}+v_2^{(1)2}}$, $\phi^{(2)}$, 
and $w^{(3)}\equiv \sqrt{w_1^{(3)2}+w_2^{(3)2}}$, as well as the values of the Jacobian of the 2-dim solutions of Tab.\ref{tab-1}.
Note that in contrast to the 1-dim case the Jacobian is not positive definite.
\begin{table}
\centering
\caption{The spin-0, spin-1,spin-2, and spin-3 contents $\phi^{(0)}$, $v^{(1)}$, $\phi^{(2)}$, and $w^{(3)}$, as well as the values of the Jacobian 
of the 2-dim solutions of Tab.\ref{tab-1}} 
\label{tab-2}       
\begin{tabular}{llllllllll}
\hline
     &$\phi^{(0)}$ & $\phi^{(2)}$ & $x_3$ & $v^{(1)}$  & $w^{(3)}$ & $\sqrt{\phi^{(0)2}+\phi^{(2)2}}$ & $\sqrt{x_3^2+v^{(1)2}+w^{(3)2}}$ & $\cal{J}$\\
\hline
1    &0.74 &0.40 &0.24 & 0.39    &0.28 &\ \ \ \ \ \ 0.84 & \ \ \ \ \ \ \ \ \ \ \ 0.54 & -0.61\\
\hline
2   &0.88  & 0.28&0.31 & 0.22    &0.10&\ \ \ \ \ \ 0.92 &\ \ \ \ \ \ \ \ \ \ \ 0.39 & 1.70\\
\hline
3   &0.66  &0.09&0.17  & 0.59    &0.42&\ \ \ \ \ \ 0.66 & \ \ \ \ \ \ \ \ \ \ \  0.75 & -0.17\\
\hline
4   &0.55  &0.08 &0.43 & 0.61    &0.37&\ \ \ \ \ \ 0.56 & \ \ \ \ \ \ \ \ \ \ \  0.83 & 0.20\\
\hline
5   &0.58  &0.47&0.57  & 0.20    & 0.28&\ \ \ \ \ \ 0.75 & \ \ \ \ \ \ \ \ \ \ \ 0.67 & 0.14\\
\hline
6   &0.41  &0.25&0.34  & 0.78   &0.19&\ \ \ \ \ \ 0.48 & \ \ \ \ \ \ \ \ \ \ \ 0.88 & -0.33\\
\hline
\end{tabular}
\end{table}
Restricting to a fundamental domain
\begin{eqnarray}
\int_{-\infty}^{+\infty} \prod_{a,b=1}^8 d A_{a1} d A_{b2}
&\rightarrow & \int_0^{2\pi}\!\!\!\!\!\! d\gamma\int_0^{\infty}\!\!\!\! r^{15} dr \int_{0<\hat{\phi}_1<\hat{\phi}_2<1}\!\!\!\!\!\!\!\!\!\!\!\!\!\!\!\!\!\!\! d\hat{\phi}_1 d\hat{\phi}_2(\hat{\phi}_2-\hat{\phi}_1) 
\int_{\rm fund.\ domain}\!\!\!\!\!\!\!\!\!\!\!\!\!\!\!\! d\alpha_1\,  d\alpha_2\, d\alpha_3\, d\alpha_4
\ {\cal J}~,
\nonumber
\end{eqnarray}
in terms of the compact variables $\hat{\phi}_i\equiv \phi_i/r$ (i=1,2) and  $\alpha_k$ (k=1,..4). Due to the difficulty of the FP-determinant $ {\cal J}$, I have, however, not yet found a satisfactory description of the fundamental domain.

\subsection{Three spatial dimensions}

Also for the general case of three dimensions, I have found several solutions of the Equ.(\ref{SAequs}) numerically for a randomly generated $A$,
see Tab.\ref{tab-3}.  Differently to the 2-dim case and the SU(2) case in 3-dim we can here achieve here  by a rotation to the
to the intrinsic systemonly either $0<\phi_1<\phi_2<\phi_3$ (solutions 1 and 2) or $\phi_1>\phi_2>\phi_3>0$ (solutions 3 and 4).
Tab.\ref{tab-4} shows the spin-0, spin-1, spin-2, and spin-3 contents\footnote{%
Note that we can represent the intrinsic variables in terms  of spin-0 fields
 $\phi^{(0)}\equiv (\phi_1+\phi_2+\phi_2)/\sqrt{3}$, spin-2 fields $\phi^{(2)}_0\equiv  \sqrt{2/3}(\phi_3-(\phi_1+\phi_2)/2)$ and $\phi^{(2)}_2\equiv (\phi_2-\phi_1)/\sqrt{2}$, the spin-1 fields 
$v_i^{(1)}\equiv\left(\sqrt{15}x_i-\sqrt{10}y_i\right)/5$, and the 7 spin-3 components $q$, $w_i$ and  $\overline{w}_i\equiv\left(\sqrt{10}x_i+\sqrt{15}y_i\right)/5$.}
$\phi^{(0)}$, $v^{(1)}\equiv \sqrt{v_i^{(1)2}}$, $\phi^{(2)}\equiv \sqrt{\phi^{(2)2}_0+\phi^{(2)2}_2}$, 
and $w^{(3)}\equiv \sqrt{w_i^{2}+\overline{w}_i^{2}+q^2}$, as well as the values of the Jacobian of the 2-dim solutions of Tab.\ref{tab-3}.
But to write the corresponding unconstrained integral over a fundamental domain
\begin{eqnarray}
\int_{-\infty}^{+\infty} \prod_{a,b,c=1}^8\!\!\! d A_{a1} d A_{b2} d A_{c3}
&\rightarrow&\int\!\! d\alpha \sin\beta\, d\beta\, d\gamma\int_0^{\infty}\!\!\!\! r^{23} dr\!\! \int_{0<\hat{\phi}_1<\hat{\phi}_2<\hat{\phi}_3<1}\!\!\!\!\!\!\!\!\!\!\!\!\!\!\!\!\!\!\!\!\!\!\!\!\!\!\!\! d\hat{\phi}_1 d\hat{\phi}_2 d\hat{\phi}_3\prod_{i<j} (\hat{\phi}_i-\hat{\phi}_j) \!\!
\int_{\rm fund.\ domain}\!\!\!\!\!\!\!\!\!\!\!\!\!\!\!\!\!\!\!\!\!\!\! d\alpha_1 ... d\alpha_{9}
\ {\cal J}~,
\nonumber
\end{eqnarray}
in terms of the compact variables $\hat{\phi}_i\equiv \phi_i/r$ and  $\alpha_k$ is a difficult, but I think solvable, future task.

\begin{table}
\centering
\caption{Several 3-dim solutions (Gribov-copies) for one randomly generated $A$, written in terms of intrinsic coordinates (the three Euler angles describing
the orientation of intrinsic frame are not shown here).}
\label{tab-3}
\begin{tabular}{lllllllllllllllllll}
\hline
&$\phi_1$&$\phi_2$&$\phi_3$&$ x_1$&$x_2$&$x_3$&$y_1$&$y_2$&$y_3$&$w_1$&$w_2$&$w_3$&$q$\\
\hline
1   &0.32\! &0.47\! &0.73\! &0.13\! &0.10\! &\!\!\! -0.05\! &\!\!\!-0.02\! &0.17\! &\!\!\! -0.05\! & 0.08\! &\!\!\! -0.26\! & 0.07\! &\!\!\! -0.003\!\! \\
\hline
2   &0.08\! &0.36\! &0.75\! &\!\!\! -0.07\! &0.02\! &\!\!\! -0.22\! &0.19\! &0.33\! &\!\!\! -0.07\! & 0.06\! &\!\!\! -0.24\! &0.15\! &0.13\!  \\
\hline
3    &0.73\! &0.40\! &0.27\!&\!\!\! -0.07\! &0.04\! &\!\!\! -0.19\! &\!\!\! -0.06\! &0.24\! & 0.13\! &\!\!\! -0.06\! &0.32\! &0.01\! & 0.06\! \\
\hline 
4  &0.66\! & 0.33\! & 0.02\! & 0.42\! & 0.43\! &\!\!\! <\! 0.001\! & 0.09\! & 0.20\! &\!\!\! -0.08\! &\!\!\! -0.16\! & 0.05\! & 0.05\! & 0.08\! \\
\hline
\end{tabular}
\end{table}
\begin{table}
\centering
\caption{Spin-0, spin-1, spin-2 and spin-3 contents $\phi^{(0)}$,$v^{(1)}$,$\phi^{(2)}$ and $w^{(3)}$ and values of the Jacobian $\cal{J}$ for the 3-dim solutions of Tab. \ref{tab-3}.}
\label{tab-4}
\begin{tabular}{llllllllllllllllllllll}
\hline
&$\phi^{(0)}$&$\phi^{(2)}$&$v^{(1)}$&$w^{(3)}$&$\!\!\!\sqrt{\phi^{(0)2}+\phi^{(2)2}}$&$\!\!\!\sqrt{v^{(1)2}+w^{(3)2}}$
&$\prod_{i>j}(\phi_i-\phi_j)$&\ $\cal{J} $\\
\hline
1     &0.88 &0.29 &0.12 &0.35 &\ \ 0.93 &\ \ 0.37   &\ \ \ \   0.016&\ 2.800\\
\hline
2     &0.69 &0.48 &0.29 &0.47 &\ \ 0.84 &\ \ 0.55   &\ \ \ \  0.074 & -0.254\\
\hline
3     &0.81 &0.34 &0.26 &0.40 &\ \ 0.88 &\ \ 0.48   &\ \ \ -0.020&\  0.739\\
\hline
4     &0.58 &0.45 &0.34 & 0.58&\ \  0.74 &\ \ 0.67    &\ \ \  -0.065& -0.168\\
\hline
\end{tabular}
\end{table}

\section{Conclusions}
Using a canonical transformation of the dynamical variables in the form of a generalised polar decomposition into
gauge-rotation and gaug-invariant parts, which Abelianises the non-Abelian Gauss-law constraints
to be implemented, a Hamiltonian formulation of  QCD in terms of gauge invariant dynamical variables can be achieved.
The exact implementation of the Gauss laws reduces the colored spin-1 gluons and spin-1/2 quarks to unconstrained colorless
spin-0, spin-1, spin-2 and spin-3 glueball fields and colorless Rarita-Schwinger fields respectively.
The obtained physical Hamiltonian admits a systematic strong-coupling expansion in powers of $\lambda=g^{-2/3}$,
equivalent to an expansion in the number of spatial derivatives.
The leading-order term in this expansion corresponds to non-interacting hybrid-glueballs, whose
low-lying masses can be calculated with high accuracy (at the moment only for the unphysical,
but technically much simpler 2-color case) by solving the Schr\"odinger-equation of Dirac-Yang-Mills quantum mechanics of spatially constant fields.
Higher-order terms in $\lambda$  lead to interactions between the hybrid-glueballs and can be taken into account systematically,
using perturbation theory in $\lambda$, allowing for the study of the difficult questions of Lorentz invariance and coupling constant renormalisation in the IR. 
First numerical investigations of the algebraic equations (\ref{SAequs}) 
give strong support for the existence of the symmetric gauge for low-energy QCD. For carrying out precise calculations as in the 2-color case,
a satisfactory description of the fundamental domain is necessary.


\end{document}